\documentclass[twocolumn,letter]{jpsj2}
\usepackage{times}
\def\d{{\rm d}}

\def\kf{k_{\rm F}}
\def\et{{\it et al}.}
\def\lsim{\lower -0.3ex \hbox{$<$} \kern -0.75em \lower 0.7ex \hbox{$\sim$}}
\def\jo #1#2#3#4{#1 {\bf #2} (#3) #4}

\def\PRB{Phys.\ Rev.\ B}
\def\PRL{Phys.\ Rev.\ Lett.}

\def\JPIF{J.\ Phys.\ I\ France}

\def\JPSJ{J.\ Phys.\ Soc.\ Jpn.}

\def\SOV{Sov.\ Phys.\ JETP}

\def\EPL{Europhys.\ Lett}

\def\JPIVF{J.\ Phys.\ IV\ France}
\def\CR{Chem.\ Rev.}

\title{%
Finite-Temperature Charge-Ordering Transition and Fluctuation Effects 
in Quasi-One-Dimensional Electron Systems at Quarter Filling  
}

\author{%
Hideo \textsc{Yoshioka}%
  $^1$\thanks{E-mail address: h-yoshi@cc.nara-wu.jp},
Masahisa \textsc{Tsuchiizu}%
  $^2$\thanks{E-mail address: tsuchiiz@slab.phys.nagoya-u.ac.jp}, and
Hitoshi \textsc{Seo}%
  $^3$\thanks{E-mail address: seo@post.kek.jp} 
}

\inst{%
$^1$Department of Physics, Nara Women's University, Nara 630-8506 \\
$^2$Department of Physics, Nagoya University, Nagoya 464-8602 \\
$^3$Non-Equilibrium Dynamics Project, ERATO-JST, 
c/o KEK, Tsukuba 305-0801
}

\recdate{\today}

\abst{%
Finite-temperature charge-ordering phase transition 
in quasi one-dimensional (1D) molecular conductors is investigated theoretically,  
based on a quasi 1D extended Hubbard model at quarter filling with  interchain
Coulomb repulsion $V_\perp$. The interchain term is treated 
within mean-field approximation 
whereas the 1D fluctuations in the chains are fully taken into account by the bosonization theory.
Three regions are found depending on how 
the charge ordered state appears at finite
temperature  when $V_\perp$ is introduced: 
 (i) weak-coupling region where the system transforms from a metal to
 a charge ordered insulator 
 with finite transition temperature at a finite critical value of $V_\perp$, 
 (ii) an intermediate region where this transition occurs by infinitesimal $V_\perp$ 
 due to the stability of inherent 1D fluctuation, and 
 (iii) strong-coupling region where the charge ordered state is realized
 already in the purely 1D case, of which the transition temperature 
becomes finite with  infinitesimal $V_\perp$.
Analytical formula for the $V_\perp$ dependence of the transition
 temperature 
is derived for each region. 
}

\kword{%
charge-ordering transition, quarter filling, quasi one-dimension,
extended Hubbard model}

\begin{document}
\maketitle

Charge-ordering (CO) phenomenon has been one of the central research subjects 
 in quasi one-dimensional (1D) organic 2:1 salts such as 
 (DCNQI)$_2X$\cite{Hiraki1998PRL,Nogami1999JPhys,Itou2004PRL}
 and (TMTTF)$_2X$\cite{Chow2000PRL,Monceau2001PRL,Nakamura2003JPSJ} 
 ($X$: monovalent counter ion). 
Some members of  them having a \textit{quarter-filled} electron or hole band 
 exhibit insulating behavior due to CO. 
For example, in (DI-DCNQI)$_2$Ag,
 an NMR measurement clarified the CO phase 
 transition at $T_{\rm CO} =220$ K\cite{Hiraki1998PRL}, 
 above which non-metallic behavior is already seen 
 even from room temperature interpreted as due to CO fluctuation\cite{Itou2004PRL}. 
In (TMTTF)$_2X$, 
CO is found by NMR as well\cite{Chow2000PRL,Nakamura2003JPSJ} 
 and the dielectric constant shows a divergence toward $T_{\rm CO}$, 
where $T_{\rm CO}=70\sim 160$ K depending on $X$\cite{Monceau2001PRL}. 

Theoretical studies showed that a minimal model for describing such CO
 is the 1D quarter-filled extended Hubbard model (EHM) 
 with not only the on-site Coulomb repulsion $U$ 
 but also the intersite repulsion
 $V$
\cite{Seo04CR}. 
This model has been intensively studied and 
the \textit{ground-state} properties have been revealed.
Using different numerical techniques, 
  the $T=0$ phase diagram on the ($U$,$V$) plane has been
 obtained with high accuracy.\cite{Numerical}
The CO insulator (COI) appears 
in the large ($U$,$V$) region.   
Analytical approaches based on the bosonization theory 
 clarified the mechanism of the transition 
 from a Tomonaga-Luttinger liquid (TLL) metallic phase 
 to the COI.\cite{Schulz94Book,Giamarchi97PhysicaB,Yonemitsu97PRB,%
     Yoshioka00JPSJ}

It is useful to consider the TLL 
 parameter $K_\rho$ depending on ($U$,$V$) in the metallic phase of this
 1D EHM.  
It takes $K_\rho > 1/4$ and approaches to $K_\rho=1/4$ at the boundary 
between the TLL and  
the COI.  
Based on the TLL theory, 
  CO fluctuation is known to develop as
  $\chi_{\mathrm{CO}}(T) \propto T^{4K_\rho-2}$, 
  which implies that for $1/2> K_\rho >1/4$ 
it diverges at low temperatures. 
This divergence does not immediately point to the  appearance of long
range order of CO in the purely 1D systems.  
The COI
is achieved for larger ($U$,$V$), only at $T = 0$. 
These features are schematically illustrated in Fig.\ 1 (a). 
When $U$ is fixed at a large value (e.g., $U\gtrsim 4t$; see Fig.~\ref{fig2} later) and 
$V$ is increased, $K_\rho$ decreases, and 
the system transforms through three regions 
bounded by $V=V_{f}^{\mathrm{1D}}$ and $V=V_{c}^{\mathrm{1D}}$ 
defined by the values where $K_\rho=1/2$ and $1/4$, respectively. 
The system is the TLL for 
(i) $V < V_{f}^{\mathrm{1D}}$ and  
(ii) $V_{f}^{\mathrm{1D}} < V < V_{c}^{\mathrm{1D}}$ 
where 
in the latter CO fluctuation develops at low-$T$, 
and it is the COI for (iii) $V_{c}^{\mathrm{1D}} < V$.
\begin{figure}[b]
\begin{center}\leavevmode
\includegraphics[width=8.5cm]{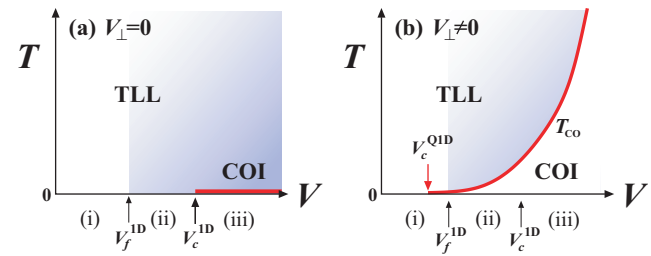}
\end{center}
\vspace{-1em}
\caption{
Schematic illustration of the finite-temperature phase
  diagram for purely 1D case (a) and quasi-1D case with finite $V_\perp$
 (b), where the CO fluctuation develops in the shaded regions.
Here $V_c^{\mathrm{Q1D}}$ express the critical value of $V$ for 
the COI appearing in the quasi-1D case. 
The quantities $V_c^{\mathrm{1D}}$ and $V_f^{\mathrm{1D}}$, and the
 regions, (i), (ii) and (iii) are defined in the text. 
}
\end{figure}

In order to examine the \textit{finite-temperature} CO transition in 
 the quasi 1D materials, in addition to such 1D fluctuation 
 we need to take into account the dimensionality effect 
by the interchain coupling. 
In these systems two types of interchain coupling exist in the electronic sector; 
  one is the single-particle interchain hopping 
 and the other is the Coulomb interaction 
  between electrons in the different chains. 
In this letter, we consider the latter and discuss the critical
 temperature of CO, $T_{\mathrm{CO}}$, 
by treating the interchain interaction within the mean-field approximation, 
whereas taking full account of the 1D fluctuations in the chains by the
 bosonization theory.  
We will find distinct behavior for the three regions defined above : 
in region (i) the system transforms from TLL metal to COI 
 with finite $T_{\mathrm{CO}}$ at a finite critical value of $V_\perp$, 
in (ii) this transition occurs by infinitesimal $V_\perp$,  and 
 in (iii) COI in the purely 1D case always has finite $T_{\mathrm{CO}}$ when $V_\perp$ is turned on. 
A schematic phase diagram in the presence of $V_\perp$ is summarized in Fig.\ 1 (b). 

 
Our Hamiltonian is given by 
  $H_{\mathrm{Q1D}}= \sum_j H^{j}_{\mathrm{1D}}+H_\perp$.
Here $H^{j}_{\mathrm{1D}}$ represents
the $j$-th isolated extended Hubbard chain:
\begin{align}
H^{j}_{\mathrm{1D}} =&
-t \sum_{i,s} \left(c^\dagger_{i,j,s} c_{i+1,j,s}^{} + \mathrm{h.c.} \right) 
\nonumber \\ &{}
 + U \sum_{i} n_{i,j,\uparrow} \, n_{i,j,\downarrow}
+ V \sum_{i} n_{i,j} \, n_{i+1,j},
\end{align}
and $H_\perp$ expresses the interchain coupling:
\begin{equation}
H_\perp  = V_\perp \sum_{i,\langle j,j'\rangle} n_{i,j} \, n_{i,j'},
\label{eqn:Hperp}
\end{equation}
where $c^\dagger_{i,j,s}$ is the creation operator of an electron 
with spin $s = \uparrow\!\! / \!\! \downarrow$
  at the $i$-th site in the $j$-th
chain, $n_{i,j,s} = c^\dagger_{i,j,s} c_{i,j,s}-1/4$ and
  $n_{i,j}=n_{i,j,\uparrow}+n_{i,j,\downarrow}$. 
In eq.(\ref{eqn:Hperp}), $\langle j,j' \rangle$ denotes the pair of
adjacent chains and  the strength of interchain Coulomb repulsion is
expressed by $V_\perp$.  
The number of adjacent chains, $z$, 
for (DCNQI)$_2X$ and (TMTTF)$_2X$ would be 
$z=4$ and $z=2$ judging from the crystal structures, 
although their interchain networks are rather 
 complicated, as we will discuss later. 
We do not take into account the interchain hopping which should not 
  yield qualitative changes to the present analysis 
  at the temperature above the crossover energy scale due to the 
  interchain hopping.\cite{Crossover}

We treat eq. (\ref{eqn:Hperp})
 in the interchain mean-field approach\cite{Scalapino75PRB}, 
 which is known to be effective in the weak $V_\perp$ region, 
 by considering the Wigner-crystal-type CO pattern stabilized in the 1D EHM. 
We assume the CO pattern to be anti-phase between adjacent chains, 
 which is naturally expected to gain the $V_\perp$ term. 
The resulting effective 1D Hamiltonian is written as 
\begin{align}
H =& 
-t \sum_{i,s} \left(c^\dagger_{i,s} c_{i+1,s}^{} + \mathrm{h.c.} \right)
 \nonumber \\ 
&+ U \sum_{i}  n_{i,\uparrow} \, n_{i,\downarrow} 
 + V \sum_{i} n_{i} \, n_{i+1} 
\nonumber \\
&+ z V_\bot n \sum_i (-1)^i n_i  
+ z N V_\bot n^2/2,
\label{eqn:eff1D} 
\end{align}
where the chain index $j$ is suppressed and $N$ is 
the total number of sites in a chain. 
The amplitude of CO is written as $n$. 

To obtain a qualitative insight of this model 
 we first discuss the 
   $U \to \infty$  limit, 
 where some exact results for the 1D model based on the Bethe ansatz can be used. 
In this case,
  we can neglect the spin degree of freedom which acts freely,
  and the charge degree of freedom is reduced 
  to a half-filled spinless fermion model:\cite{Ovchinnikov73Sov}
\begin{align}
H_{U\to\infty} =& 
-t \sum_{i} (d^\dagger_{i} d_{i+1}^{} + \mathrm{h.c.} )
+ V \sum_{i} \tilde n_{i} \, \tilde n_{i+1} 
\nonumber \\ & 
+ z V_\bot n \sum_i (-1)^i \tilde n_i
+ z N V_\bot n^2/2,
\label{eq:H_Uinf} 
\end{align}
where the creation operator of the spinless fermion is expressed by
$d_i^\dagger$ and  $\tilde n_i \equiv d_{i}^\dagger \, d_{i}^{}-1/2$. 
This model is equivalent to the $S=1/2$ XXZ spin-chain coupled by the
  interchain exchange interaction $J_\perp(= V_\perp)$
  treated in the interchain mean-field approach.\cite{Inagaki88JPSJ,Giamarchi2004book}    
Several authors have performed 
  such approach to the Heisenberg limit (i.e., $V=2t$).
\cite{Schulz96PRL}
Bhaseen and Tsvelik discussed a nontrivial spectrum of eq. (\ref{eq:H_Uinf})
   in a parameter region $V \ge 2t$,\cite{cond-mat/0409602}
   where the system is the COI and has a finite excitation gap 
   even for $V_\bot = 0$.\cite{Ovchinnikov73Sov}
   We restrict ourselves to a parameter region $0< V \le 2t$, where the 
   system is known to be metallic if $V_\bot = 0$, and show that infinitesimaly
   small $V_\bot$ makes the system insulating and gives rise to finite
   $T_{\mathrm {CO}}$.

We apply the bosonization method to eq. (\ref{eq:H_Uinf}). 
The density operator is given by
${\tilde n}_i  = 
  a(\sqrt{2}\pi)^{-1} \partial_x \theta(x)
-\pi^{-1} (-1)^i \sin \sqrt{2} \theta(x)$ with $x=ia$
  and $a$ being the lattice constant.
This parameter region $0 < V \le 2t$ corresponds to region (ii) with $1/2>K_\rho \ge 1/4$ defined above, 
since it is known that in the 1D Hubbard model at $U \to \infty$ ($V=0$),
 $K_\rho$ takes a universal value $K_\rho \to 1/2$\cite{Schulz90PRL}, 
 and at $V=2t$, the CO transition indicates $K_\rho=1/4$.

In the bosonized form of the Hamiltonian, there appears
  a nonlinear term, $V/(2\pi^2 a) \cos 2\sqrt{2}\theta$,
 which becomes irrelevant but has an effect to renormalize
  the velocity, the TLL parameter and 
  the prefactor of the bosonized density operator.
By taking into account these effects properly,
  the density operator is given by
$\tilde n_i = 
a(\sqrt{2}\pi)^{-1} \partial_x \theta(x)
-c \pi^{-1} (-1)^j \sin \sqrt{2} \theta(x)$,
  where $c$ is a nonuniversal constant depending on $V/t$
 and its exact value has been suggested.\cite{Lukyanov99PRB}
We will use the renormalized  velocity and the TLL parameter,  
  and neglect the  $\cos 2\sqrt{2}\theta$ term.
After this procedure, 
we switch on the $V_\perp$ term, and then 
the effective Hamiltonian density reads
\begin{align}
\mathcal{H}^{\mathrm{eff}}_{U\to\infty}=&
\frac{v}{4\pi} \left[
\frac{1}{2K_\rho} (\partial_x \theta)^2 + 2K_\rho  (\partial_x \phi)^2
\right]
\nonumber \\ & {}
- \frac{g_\perp}{2\pi^2 a^2} \sin \sqrt{2} \theta
+\frac{z}{2a} V_\perp n^2
,
\label{eq:Hboson2}
\end{align}
where 
$v=\pi t a \sqrt{1-(V/2t)^2}   /\arccos(V/2t)$,
$K_\rho =[4-(4/\pi)  \cos^{-1} (V/2t)]^{-1}$,
  and $g_\perp = 2\pi c z V_\perp a n$.
The TLL parameter $K_\rho$ determines 
the nonuniversal constant $c=c(K_\rho)$. 

Since the $g_\perp$ term in eq.\ (\ref{eq:Hboson2})
  is a relevant perturbation, when $n~\ne~0$ the system is not the TLL but
  has an excitation gap, 
\cite{Zamolodchikov}
\begin{equation}
m(n)
=
\frac{v}{a} \left[
\frac{g_\perp}{4\pi^2 v \, \kappa\bigl(\frac{K_\rho}{1-K_\rho}\bigr)}
\right]^{1/(2-2K_\rho)},
\label{eq:m}
\end{equation}
where
$\kappa(p) \equiv \pi^{-1}(\pi/4)^{1/(p+1)}\Gamma\bigl( p/(p+1) \bigr)$
$\Gamma^{-1}\bigl(1/(p+1)\bigr)$ $\bigl[ \Gamma\left( (p+1)/2 \right)$
$\Gamma^{-1} \left( p/2 \right)\bigr]^{2/(p+1)}.$
The ground-state energy per site is given by\cite{Zamolodchikov}
\begin{equation}
E(n)=
- \frac{a}{4v}  m^2(n) \tan \left( \frac{\pi K_\rho}{2-2K_\rho} \right)
+zV_\perp n^2/2 .
\label{eq:e} 
\end{equation}
The CO amplitude $n$ is determined from the condition to minimize 
eq. (\ref{eq:e}). 
The optimized magnitude $n$ is given by
\begin{align}
n =&
\frac{1}{2}\left[
\frac{\tan \bigl( \pi K_\rho/(2-2K_\rho) \bigr)}{2-2K_\rho} 
\right]^{\frac{1-K_\rho}{1-2K_\rho}}
\nonumber \\ & {} \times
\left[\frac{c(K_\rho)}
           {2\pi \kappa\bigl(\frac{K_\rho}{1-K_\rho}\bigr)}\right]
 ^{\frac{1}{1-2K_\rho}} 
\left( \frac{z V_\perp a}{2v} \right)^{K_\rho/(1-2K_\rho)} .
\label{eq:n}
\end{align}
Inserting this expression to eqs.~(\ref{eq:m}) and (\ref{eq:e}), 
 the excitation gap and the ground-state energy are given by
 $m=m(n)\propto  V_\perp^{1/(2-4K_\rho)}$
  and $E = E(n) \propto V_\perp ^{1/(1-2K_\rho)}$. 
Thus, an \textit{infinitesimal} value of interchain interaction transforms the system 
 from TLL in the 1D case to the COI 
 for \textit{all} values of $0 < V \le 2t$.  
${T_{\mathrm{CO}}}$ can also be 
  estimated in a similar way.
The free energy up to  $n^2$ is given by
\begin{align}
f(n)
=& 
-\frac{\pi T^2 }{6v} 
-\frac{b(K_\rho)}{4 \pi^2 v} 
\frac{ z^2 V_\perp^2 n^2}{(2\pi T a/v)^{2-4K_\rho}}
+ \frac{z V_\perp n^2}{2a}, 
\label{eq:free}
\end{align}
where $b(K_\rho)\equiv  c^2(K_\rho)\sin( 2\pi K_\rho )
  B^2 ( K_\rho, 1-2K_\rho )$  with
 $B(x,y)=\Gamma(x)\Gamma(y)/\Gamma(x+y)$.
From the condition where the coefficient of $n^2$ 
in eq.\  (\ref{eq:free}) vanishes,  
 $T_{\mathrm{CO}}$ is determined as 
\begin{eqnarray}
 T_{\mathrm{CO}}
=
\frac{v}{2\pi a}
\Biggl[
\frac{z b(K_\rho)}{ 2\pi^2 v}  V_\perp a
\Biggr]^{1/(2-4K_\rho)} .
\label{eq:Tco}
\end{eqnarray}
At $V=2t$, 
${T_{\mathrm{CO}}}$ is proportional to 
$V_\perp$.\cite{Inagaki88JPSJ,Giamarchi2004book}
In this case, 
the ratio of the gap at $T=0$ and the transition temperature 
becomes  $m/T_{\mathrm{CO}} |_{V= 2 t-}\approx 2.47$,
which reproduces the result obtained 
in Ref. \citen{Orignac2004}.
The formulas of 
the amplitude $n$ [eq.\ (\ref{eq:n})] and the transition temperature
  $T_{\mathrm{CO}}$ [eq.\ (\ref{eq:Tco})] 
 are valid for $V\lsim V_c^{\mathrm{1D}}$, since these
  diverge when $V\to 0$
  due to the unphysical ultraviolet divergence in 
  eqs.\ (\ref{eq:e}) and (\ref{eq:free}). 
This problem could be resolved simply by introducing an
  ultraviolet cutoff.

Now we consider the case of finite $U$. 
An effective Hamiltonian for low-energy states  ($k \approx \pm \pi/(4a)$) 
can be obtained by integrating out high-energy  states.%
\cite{Yoshioka00JPSJ}  
The Hamiltonian is separated into the charge part describing the 
CO transition and the spin part.  
The latter is essentially the same as 
the Hamiltonian of 1D isotropic Heisenberg model and leads to 
gapless excitations both in the TLL and the COI. 
The density operator can be bosonized as
\begin{align}
 n_i =& \frac{a}{\pi} \partial_x \theta_\rho(x)
+ \frac{2c_1}{\pi} \cos(2\kf x+\theta_\rho(x))\sin\theta_\sigma(x)
\nonumber \\
&+ c_2 \frac{ (-1)^i}{\pi} \cos 2\theta_\rho(x),
\end{align} 
where $\theta_{\rho}(x)$ and $\theta_\sigma(x)$ are the bosonic 
  phase variables for the charge and spin degrees of freedom.
$c_1$ and $c_2$ are nonuniversal numerical constants.
In the absence of $n$, the charge part 
is written by the following phase Hamiltonian,
\begin{align}
 {\cal H}_\rho =& \frac{v_\rho}{4 \pi} 
\left[
\frac{1}{K_{\rho0}} (\partial_x \theta_\rho)^2
         + K_{\rho0} (\partial_x \phi_\rho)^2
\right] 
+ \frac{g_{1/4}}{2 (\pi a)^2} \cos 4 \theta_\rho, 
\label{eqn:1/4no-n}
\end{align}
where $[\theta_\rho(x),-\partial_{x'}\phi_\rho(x')/(2\pi)] 
= \mathrm{i} \delta(x-x') $. 
The charge velocity  $v_\rho$, the parameter $K_{\rho0}$, and
   the magnitude of the 1/4-filled Umklapp scattering $g_{1/4}$
  are nonuniversal and its expressions obtained 
  perturbatively are given in Ref.\ 
  \citen{Yoshioka00JPSJ}. 
In the presence of the interchain mean-field $n$, the additional Umklapp 
scattering appears, as in the last two terms in eq.\ (\ref{eq:Hboson2}), 
\begin{align}
{\cal H}' &= \frac{zc_2 }{\pi a} V_\bot n
 \cos 2 \theta_\rho + \frac{z}{2a}V_\bot  n^2.
\label{eqn:1/4n}       
\end{align}
By minimizing $\langle {\cal H}_\rho + {\cal H}' \rangle$ with respect to $n$,
the amplitude  $n$ is determined as
\begin{align}
 n = - \frac{c_2}{\pi} \langle \cos 2 \theta_\rho \rangle,
\label{eqn:n} 
\end{align} 
where $\langle \cdots \rangle$ denotes the finite-$T$ expectation value 
  with respect to $\mathcal{H}_\rho+\mathcal{H}'$.
We note that the parameter $c_2$ is proportional to the on-site
interaction, $ c_2 \approx  {U}/{(\sqrt{2}\pi t)}$ within the perturbative treatment. 
This implies that the CO fluctuation with 2-fold periodicity 
 can emerge  due to the correlation effect.

In order to estimate the 
  second-order transition temperature, 
 the lowest-order perturbation theory 
 in $n$ is sufficient. 
By expanding the r.h.s.\ of  eq.\ (\ref{eqn:n})
 with respect to $\mathcal{H}'$,
the equation to determine $T_{\mathrm{CO}}$
  is obtained as\cite{Giamarchi2004book}
\begin{align}
 1 =&  \frac{z c_2^2}{\pi^2 a} V_\bot 
 \int \mathrm{d} x  \int_0^{1/T_{\mathrm{CO}}} \mathrm{d} \tau
\nonumber \\
& \times
 \langle  T_\tau \cos 2 \theta_\rho (x, \tau) 
\cos 2 \theta_\rho (0,0)\rangle_0|_{T=T_{\mathrm{CO}}},
\label{eq:selfconsistent0}
\end{align}
where $\langle \cdots \rangle_0$ is the expectation value 
  with respect to $\mathcal{H}_\rho$.
In the following, we employ naive renormalization group (RG) arguments to estimate $T_{\mathrm{CO}}$. 
By neglecting the anisotropy in space and time, 
the correlation function 
$\langle T_\tau \cos 2 \theta_\rho (x, \tau)
\cos 2 \theta_\rho (0,0)\rangle_0$
can be roughly calculated as
\begin{align}
 &\langle T_\tau \cos 2 \theta_\rho(x,\tau) 
\cos 2 \theta_\rho(0,0) \rangle_0 \nonumber \\ 
&\approx \frac{1}{2}
\exp \left[ - \int_0^{\ln [\mathrm{min}(r_\rho,\xi_\rho)/a]} \mathrm{d} l
  \bigl( 4 K_{\rho}(l) + 2 G_{1/4} (l) \bigr)\right],  
\label{eq:cf}
\end{align} 
where $r_\rho = \sqrt{x^2 + (v_\rho \tau)^2}$ and
 $\xi_\rho=v_\rho/(\pi T)$ is the characteristic 
  thermal coherence length.
The quantities $K_{\rho}(l)$ and $G_{1/4}(l)$ are the solution of the RG
equations, 
\begin{subequations}
\begin{align}
 \d K_{\rho}(l) / {\mathrm{d} l}  &= - 8 G_{1/4}^2(l) K_{\rho}^2(l), \\
 \d G_{1/4}(l) / {\mathrm{d} l}  &= (2 - 8  K_{\rho}(l))G_{1/4}(l),  
\end{align}%
\label{eq:RG}%
\end{subequations}
where the initial conditions are 
$K_\rho(0) = K_{\rho0}$ and $G_{1/4}(0) = g_{1/4}/(2 \pi v_\rho)$. 
These RG equations have two kinds of fixed points
  for realistic parameters.\cite{Yoshioka00JPSJ}
One is $(K_\rho(\infty),G_{1/4}(\infty))=(0,-\infty)$ and 
the  other is $(K_\rho(\infty),G_{1/4}(\infty))=(K_\rho,0)$.
For $V_\perp=0$ the former corresponds to 
the COI at $T=0$, while the latter 
  to the metallic TLL  state.
From eqs.\ (\ref{eq:selfconsistent0}) and (\ref{eq:cf}),
${T_{\rm CO}}$ 
 can be estimated by 
$F\left({b v_\rho}/{T_{\rm CO} }\right) = 1$
where
\begin{align}
F(s)
\equiv
& \frac{z c_2^2 a V_\bot}{\pi v_\rho} \int_0^{\ln s}
dy \nonumber \\
&\times \exp \left[ - \int_0^y \d l 
  \bigl( 4 K_\rho(l) + 2 G_{1/4}
 (l) -2 \bigr) \right],
\label{eq:F}
\end{align}
and $b$ is a positive numerical constant. 

Concerning  the realization of the COI at finite-$T$,  
we find three distinct regions on the ($U$,$V$) plane shown in Fig.2.
\begin{figure}[b]
\vspace{-1em}
\begin{center}\leavevmode
\includegraphics[width=7.2cm]{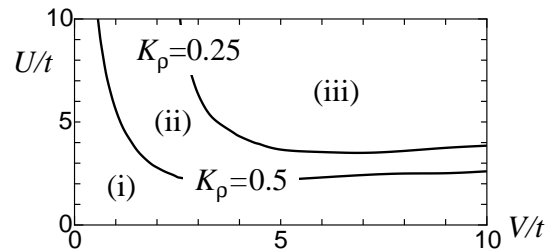}
\end{center} 
\vspace{-1em}
\caption{Three distinct regions classified by appearance of 
the COI at the finite temperature.  
}
\label{fig2}
\end{figure}
In region (iii) where the COI is already realized in the purely 1D case, 
infinitesimal $V_\bot$ makes $T_{\mathrm{CO}}$ 
finite, as is naturally expected.
In addition, there is a region (ii) with $1/2 \ge K_\rho \ge  1/4$ 
where \textit{infinitesimal} $V_\perp$ also gives rise to finite
$T_{\mathrm{CO}}$
even in the metallic region for $V_\perp=0$,  
 since 
$F(\infty)$ diverges due to the divergence of CO fluctuation while $G_{1/4}(\infty)=0$. 
This is consistent with the case of $U \to \infty$ discussed above, 
where $0<V\le2t$ correspond to region (ii) in this limit, as mentioned.
Moreover, 
even in the region (i) without divergence of CO fluctuation, 
\textit{finite} amount of $V_\bot$ makes the system COI. 
The obtained expressions for 
$T_{\mathrm{CO}}$ as a function of $V_\perp$ 
is summarized in Table. I. 
The critical value of $V_\bot$ for appearance of the COI 
in the region (i) is proportional to $4 K_\rho -2$.
We note that $T_{\mathrm{CO}} \propto \exp(-A/V_\perp)$ at $K_\rho = 1/2$, and
$T_{\mathrm{CO}} \propto V_\perp$ for $K_\rho = 1/4$, 
also consistent with the $U \to \infty$ case.  
\begin{table}[tb]
\caption{The $V_\perp$ dependence of the transition temperature
   $T_{\mathrm{CO}}$ in the respective regions,  
where $A$, $B$ and $C$ are positive constants.
Corresponding $K_\rho$ values are also listed.
In the region (i), the transition happens only for $V_\perp>A(4K_\rho-2)$. 
}
\label{t1}
\begin{tabular}{c|ll}
\hline\hline
 & $K_\rho$ & $T_{\mathrm{CO}}$ 
\\ \hline
(i) & $K_\rho\ge 1/2$ & $\propto [1-{A(4K_\rho-2)}/{V_\perp}]^{1/(4K_\rho-2)}$
\\
(ii) & $1/4 \le  K_\rho \le 1/2$  & $\propto [1-{A(4K_\rho-2)}/{V_\perp}]^{1/(4K_\rho-2)}$
\\
(iii) & not defined & $=V_\perp[1+B\ln(C/V_\perp)]$  
\\
\hline\hline
\end{tabular}
\vspace{-1.5em}
\end{table}

Now let us discuss the relevance of our results to the experiments. 
In (DI-DCNQI)$_2$Ag, 
  the non-metallic behavior above $T_{\rm CO}\simeq$ 220 K\cite{Itou2004PRL} 
  might be due to the $1/4$-filled Umklapp scattering 
  described in eq.\ (\ref{eqn:1/4no-n}). 
The peak  at $T=T_{\rm CO}$ in the derivative of the resistivity as a function of $T$ in Ref. \ref{Itou2004PRL} 
 seems to be triggered by the generation of additional Umklapp scattering $\cos 2\theta_\rho$
 in eq.\ (\ref{eqn:1/4n}) due to the appearance of the CO. 
In (TMTTF)$_2X$, the system is also insulating already above $T_\mathrm{CO}$ 
 and a change in the slope of the resistivity curve is observed.\cite{Laversanne1984} 
However, in this case the slight dimerization in the chain direction 
gives rise to another relevant $1/2$-filled Umklapp scattering\cite{Yoshioka00JPSJ}. 
Whether or not the $1/4$-filled Umklapp scattering is effective above $T_\mathrm{CO}$ 
 is difficult to judge, but the change in the slope should be due to the 
 additional Umklapp scattering by CO as discussed in this paper.

In these quasi 1D compounds, 
 the interchain Coulomb interaction is in fact considered to be fairly large. 
For example in some members of (TMTTF)$_2X$ 
 it is even estimated to be comparable to the intrachain ones
 in quantum chemistry calculations\cite{Fritsch91JPIF}. 
This is noticeable since the interchain electron hopping is one order of magnitude 
 smaller than that in the intrachain direction due to the anisotropic shape of the HOMO of the TMTTF molecule. 
However we should note also that the actual interchain networks are rather complicated. 
In the structure of (DCNQI)$_2X$, 
 it has a spiral symmetry when rounding each plaquette in the network. 
It is pointed out that this gives rise to frustration 
 for the Wigner-crystal-type CO since the periodicity of the spiral 
 does not fit the 2-fold period along the DCNQI chains\cite{Kanoda05JP4}. 
In (TMTTF)$_2X$, such frustration is more straightforwardly expected 
 since the interchain coupling is zigzag-like therefore anti-phase and in-phase 
 CO patterns would have close energy. 
These should reduce the effective interchain interaction to a smaller value 
 at a first approximation. 

In the present analysis, we considered the Wigner-crystal-type CO with
  a 2-fold periodicity only, namely, 
 the ``$4\kf$'' CDW state.
In the small $V$ region, however, the $2\kf$ CDW state having a 4-fold periodicity
  may also be stabilized, 
since its fluctuation develops as 
 $\chi_{c}^{2\kf}(T) \propto T^{K_\rho-1} $.
From the simple power counting, one finds that
    $2\kf$ fluctuation becomes dominant  if $K_\rho>1/3$.
Thus in the phase diagram where $1/3<K_\rho<1/2$, 
 competition and/or coexistence of the CO and the $2\kf$ CDW
   may be possible. 
This problem needs further investigation.

In conclusion, we investigated the finite-temperature 
 CO transition in the quasi-1D electron
 system at quarter filling coupled by the interchain Coulomb repulsion $V_\perp$,
 by using the interchain mean-field theory and the bosonization method. 
It was shown that the interchain interaction gives rise to finite transition temperature $T_{\rm CO}$, 
 with different behavior depending on the parameter in the 1D limit. 
When it is in the COI,  $V_\perp$ gives rise to 
 finite temperature phase transition. 
In the TLL phase, 
 in the parameter region where the charge-fluctuation develops, 
 i.e., $1/2 > K_\rho \ge 1/4$, 
 infinitesimally small $V_\perp$ also produces the COI with finite $T_{\rm CO}$,  
 while for $K_\rho<1/2$, at a finite critical value of $V_\perp$ 
 the system transform from the TLL to the COI. 
  
\section*{Acknowledgment}
The authors thank Y.\ Suzumura for valuable comments, and K.\ Sano and
Y.\ $\bar{\rm O}$no for giving them the numerical data of
Ref.\citen{Numerical} for writing Fig. 2. 
This work was supported by Grant-in-Aid for Scientific Research on
Priority Area of Molecular Conductors (No.16038218) and 
Grant-in-Aid  for Scientific Research (C) (No.15540343)
from MEXT.

\end{document}